\documentclass[usenatbib,referee]{biom}
\usepackage{amsmath}
\usepackage{bbm}
\usepackage{dsfont}
\usepackage{xcolor}
\usepackage{float}
\usepackage{amssymb}
\usepackage{graphicx}
\usepackage{natbib} 
\usepackage{url} 
\usepackage{hyperref}
\usepackage{algorithm} 
\usepackage{mathrsfs}
\usepackage{algpseudocode} 

\def\bSig\mathbf{\Sigma}

\title[]{Analysis of block slice samplers for Bayesian GLMMs and GAMs with linear inequality and shape constraints}

\author{Benny Ren,  
Jeffrey Morris, and
Ian Barnett \\
Department of Biostatistics, Epidemiology, and Informatics, \\ 
University of Pennsylvania, Philadelphia, U.S.A.\\}

\begin{document}


\date{}



\pagerange{\pageref{firstpage}--\pageref{lastpage}} 
\volume{}
\pubyear{}
\artmonth{}


\doi{}


\label{firstpage}


\begin{abstract}
Exponential family models, generalized linear models (GLMs), generalized linear mixed models (GLMMs) and generalized additive models (GAMs) are widely used methods in statistics. However, many scientific applications necessitate constraints be placed on model parameters such as shape and linear inequality constraints. Constrained estimation and inference of parameters remains a pervasive problem in statistics where many methods rely on modifying rigid large sample theory assumptions for inference. We propose a flexible slice sampler Gibbs algorithm for Bayesian GLMMs and GAMs with linear inequality and shape constraints. We prove our posterior samples follow a Markov chain central limit theorem (CLT) by proving uniform ergodicity of our Markov chain and existence of the a moment generating function for our posterior distributions. We use our CLT results to derive joint bands and multiplicity adjusted Bayesian inference for nonparametric functional effects. Our rigorous CLT results address a shortcoming in the literature by obtaining valid estimation and inference on constrained parameters in finite sample settings. Our algorithmic and proof techniques are adaptable to a myriad of important statistical modeling problems. We apply our Bayesian GAM to a real data analysis example involving proportional odds regression for concussion recovery in children with shape constraints and smoothed nonparametric effects. We obtain multiplicity adjusted inference on monotonic nonparametric time effect to elucidate recovery trends in children as a function of time.
\end{abstract}

%



\maketitle


%

\section{Introduction}
\label{intro}
Many scientific domains use statistical models with constraints on parameters which often results in difficult estimation and inference problems. For example, in genetics, simplex constraints are used to account for compositional data \citep{wang2019bulk,lu2019generalized}. In risk and survival analysis, a monotonic baseline function is used account for cumulating risk over time \citep{cox1972regression}. We focus on the survival context, which is further compounded by the fact that the monotonic function is unknown. To address nonparametric regression, we propose a slice sampling Gibbs algorithm that is applicable to a broad class of generalized linear mixed models (GLMMs) and generalized additive models (GAMs) with linear inequality and shape constraints \citep{hastie2017generalized}. Our algorithm is flexible and customizable to many different settings, as well as being computationally and algebraically tractable.

In event time modeling, two popular classes of models: Cox proportional hazards (PH) and semiparametric proportional odds (PO) models, are often used to study censored outcomes \citep{rossini1996semiparametric,shen1998propotional,murphy1997maximum}. These two methods share a common structure of a monotonic increasing nonparametric baseline function of time in their regression equation. In the case of the Cox PH, the partial likelihood can be used to obtain consistent estimates of regression parameters without directly modeling the baseline hazard \citep{cox1975partial}. However, when either the baseline odds or baseline hazards is of interest, rigorous estimation and inference frameworks are necessary to study these two nonparametric functions \citep{zeng2007maximum}.

In semiparametric PO regression, we observe current status (whether a failure has occurred) at a monitoring time and a set of covariates related to failure status. From this information, we know whether a failure occurred before a monitoring time (failure time is before the monitoring time) or is censored (failure time is after the monitoring time). \cite{rossini1996semiparametric,huang1995maximum} showed that the likelihood can be simplified with an independence assumption, failure time and covariates are independent of monitoring time, leading to an ancillary statistic being removed from the likelihood during estimation. The resulting likelihood is equivalent to a logistic regression likelihood with an unknown monotonic baseline function of monitoring time. The baseline function can be dealt with using monotonic regression and be replaced by a linear combination basis functions, resulting a potentially high-dimension logistic regression \citep{hothorn2018most,shen1998propotional,hanson2007bayesian,rossini1996semiparametric}. As noted in \cite{lin2010semiparametric,wang2011semiparametric}, shrinkage priors can be used to address non-descriptive basis functions. \cite{ramsay1988monotone} outlined the use of the I-spline system in conjunction with constrained optimization for monotonic regression. Similar constraints can be used to enforce convexity and other shapes into nonparametric effects \citep{meyer2015bayesian,ghosal2023shape}. Incorporating basis functions into the data matrix, our Gibbs sampler obtains Bayesian inference of regression coefficients and the monotonic baseline function, while being able to incorporate information through a multivariate Gaussian prior and random effects \citep{wand2008semiparametric,vallejos2017bayesian,polson2013bayesian}. 

Under a Markov chain Monte Carlo (MCMC) framework, uniform ergodicity is one of the prerequisite condition for central limit theorem (CLT) inference for MCMC estimators. In addition to uniform ergodicity, finite second moments of the posterior distribution are needed to ensure CLT properties of posterior MCMC samples \citep{roberts2004general,jones2004markov}. We prove uniform ergodicity of posterior MCMC samples and the existence of a moment generating function (MGF) for our posterior distributions to obtain Markov chain CLT results. We expand on our CLT results and propose joint bands, multiplicity adjusted inference for nonparametric effects and monotonic baseline functions \citep{ruppert2003semiparametric,lee2018bayesian,meyer2015bayesian,morris2015functional}.

Constrained model estimation and inference is a difficult problem in statistical modeling. Traditional convex optimization based algorithms such as \cite{lu2019generalized}, uses a complicated descent algorithm paired with a de-biased covariance for estimation and large sample theory inference when dealing with constraints. Furthermore, use of large sample theory may not be valid in modest sample size settings such as causal inference and clinical trials. Our Bayesian MCMC approach addresses a shortcoming in the literature by simultaneously obtaining estimation and inference for finite sample settings. Our slice sampler is well suited for estimation and inference, due to its CLT properties for the posterior samples while being able to ensure linear inequality and shape constraints. Furthermore, our proof approach and Gibbs algorithm can be modified and recycled to handle similarly parameterized problems such as Bayesian variable selection and hierarchical models. Given the theoretical properties and computationally succinct formulation of our Gibbs algorithm, we believe that our slice sampler is an attractive method for numerous statistical modeling problems

\section{Methods}
\label{methods}

\subsection{Proportional odds regression with current status data and nonparametric effects}
Using the derivation of the proportional odds model from \cite{shen1998propotional,rossini1996semiparametric,huang1995maximum}, we collect censoring time $T_i$ for each subject $i\in\{1,\dots,N\}$. We observe data: $W_i=(T_i, Y_i, \mathbf{x}_i) \in \mathbb{R}^{+} \times\{0,1\} \times \mathbb{R}^{p}$ where $Y_i=\mathbb{I}\left(T^*_i \leq T_i\right)$, indicating whether event, denoted by event time $T^*_i$, has occurred or not, i.e. the current status. We do not observed the true event time $T^*_i$, but know the current status at time $T_i$. Here $p$ covariates are given as $\mathbf{x}_i$. The semiparametric proportional odds model is defined as is defined as
$$
\mathbb{E}(y_i \mid t_i, \mathbf{x}_i)= \mathrm{Pr}(y_i \mid t_i, \mathbf{x}_i) =\frac{\exp \left(\alpha(t_i) + \mathbf{x}_i^\top \boldsymbol{\beta} \right)}{1+\exp \left(\alpha(t_i) + \mathbf{x}_i^\top \boldsymbol{\beta} \right)} .
$$
Using Bayes' rule, we have 
$$
\mathrm{Pr}(W_i=w_i | \alpha, \boldsymbol{\beta}) = \mathrm{Pr}(y_i \mid t_i, \mathbf{x}_i, \alpha, \boldsymbol{\beta})\mathrm{Pr}(t_i, \mathbf{x}_i| \alpha, \boldsymbol{\beta}) = \frac{\exp \left(y_i \left(\alpha(t_i)+\mathbf{x}_i^\top \boldsymbol{\beta} \right)\right)}{1+\exp \left(\alpha(t_i)+\mathbf{x}_i^\top \boldsymbol{\beta} \right)} h(t_i, \mathbf{x}_i)
$$
where $\mathbf{x}_i, Y_i$ is assumed independent of $T_i$ and $h(t_i, \mathbf{x}_i)$ is the joint density of $(Y_i, \mathbf{x}_i)$ which does not depend on $(\alpha(t), \boldsymbol{\beta})$. Therefore $h(t_i, \mathbf{x}_i)$ is an ancillary statistic and can be omitted from the estimation. As a result, we obtain $\operatorname{logit} ( F(t_i \mid \mathbf{x}_i ) )=\alpha(t_i)+\mathbf{x}_i^{\top} \boldsymbol{\beta}$
and 
$$
\frac{F(t_i \mid \mathbf{x}_i)}{1-F(t_i \mid \mathbf{x}_i)}=\frac{F_{0}(t_i)}{1-F_{0}(t_i)} \exp \left(\mathbf{x}_i^{\top} \boldsymbol{\beta} \right), \quad \alpha(t_i) = \log \left( \frac{F_{0}(t_i)}{1-F_{0}(t_i)} \right) .
$$
The logit function is monotonic on $(0,1)$; CDF $F_{0}(t_i)$ is monotonic and unknown; and $\alpha(t)$ is monotonic. Thus we have the following likelihood contribution for subject $i$, 
\begin{equation} \label{eq0}
\mathrm{Pr} (W_i=w_i | \alpha, \boldsymbol{\beta}) \propto \left[ \frac{\exp \left( \alpha(t_i)+\mathbf{x}_i^{\top} \boldsymbol{\beta} \right)}{1+\exp \left( \alpha(t_i)+\mathbf{x}_i^{\top} \boldsymbol{\beta} \right)} \right]^{y_i} \left[ \frac{1}{1+\exp \left( \alpha(t_i)+\mathbf{x}_i^{\top} \boldsymbol{\beta} \right)} \right]^{1-y_i}.
\end{equation}
We can also write a general regression model with basis functions as $ \mathbf{m}_i^\top \boldsymbol{\eta} = \mathbf{x}_i^{\top} \boldsymbol{\beta} + \alpha(t_i) + \mathcal{S}(\mathcal{X}_i)$, where $\mathcal{S}(\mathcal{X}_i)$ is a nonparametric covariate effect.

\subsubsection{Monotonic regression with constrained coefficients}
\label{monotone}
Using I-splines ${I}_{m}(t)$, to construct a monotonic semiparametric regression using $M-2$ knots, we obtain the following regression
$$
\alpha(t) = \sum_{m=1}^{M+2} u_{\alpha,m} {I}_{m}(t)
$$
with $\mathbf{u}_\alpha$ as basis coefficients and can be expressed as $\mathbf{Z}_{\alpha} \mathbf{u}_{\alpha}$ in matrix form and the intercept $\beta_0$ being estimated without constraints \citep{ramsay1988monotone,meyer2008inference}. Note that constraints $\mathbf{u}_\alpha \geq \mathbf{0}$, guarantee a monotonic $\alpha(t)$ which can be achieved in a Gibbs sampler by sampling from a half-normal distribution. In line with Bayesian methodology for PO models, we impose shrinkage prior $\mathbf{u}_{\alpha} \sim \mathrm{N}(\textbf{0}, \tau^{-1}_{\alpha} \mathbf{I}_{M+2})$, on I-spline coefficients \citep{wang2011semiparametric}. 

\subsubsection{Semiparametric regression with O'Sullivan penalized B-splines}
\label{gam}
We may represent nonparametric effects $S (\mathcal{X})$, using cubic B-splines with $M$ number of knots
\begin{equation} \label{eq1}
S (\mathcal{X}) = \sum_{m=1}^{M+4} b_{m} \mathcal{B}_{m}(\mathcal{X}) + e_i
\end{equation}
with $b_{m}$ as basis coefficients and $e_{i} \sim \mathrm{N}\left(0, \sigma^2_{\mathcal{B}} \right)$. For each basis, $b_{m}$ are B-spline coefficients and $\mathcal{B}_{m}(\mathcal{X}), m=1, \ldots, M+4$ are basis functions defined by the knots $\psi_{1}, \ldots, \psi_{M+8}$ where,
$$
\begin{aligned} L &=\psi_{1}=\psi_{2}=\psi_{3}=\psi_{4}<\psi_{5}<\cdots<\psi_{M+4}=\psi_{M+5} \\ &=\psi_{M+6}=\psi_{M+7}=\psi_{M+8}= U \end{aligned}
$$
and $L$ and $U$ are boundary knots \citep{hastie2009elements}. 

Writing \eqref{eq1} in matrix form, we get $\boldsymbol{S} = \boldsymbol{\mathcal{B}} \boldsymbol{b}+\boldsymbol{e}
$ where $\boldsymbol{S}=\left[ S_1, \ldots, S_N \right]^{\top}$, $\boldsymbol{\mathcal{B}}$ is the $N \times(M+4)$ B-spline matrix,  $\boldsymbol{b} =\left[ b_1, \ldots, b_{M+4} \right]^{\top}$ and $\boldsymbol{e}=\left[ e_{1}, \ldots, e_{N} \right]^{\top} \sim$
$\mathrm{N}\left(\mathbf{0}, \sigma^2_{\mathcal{B}} \mathbf{I}_{N}\right)$. The O'Sullivan penalize B-splines (O-splines), defines a second order penalty term, curvature smoothness penalty, $\lambda_{\mathcal{B}} \int \left\{S^{{\prime \prime}}(\mathcal{X})\right\}^{2} d \mathcal{X}$, which is reasonable and often desirable property in semiparametric regression \citep{o1986statistical}. As noted in \cite{wand2008semiparametric}, this penalty is equivalent to assuming a prior distribution on the coefficients to be $\boldsymbol{b} \sim \mathrm{N}\left(\mathbf{0}, {q}^{-1}_{\mathcal{B}} \boldsymbol{\Lambda} \right)$, $\left[ \boldsymbol{\Lambda} \right]_{mm^{\prime}} = \int {\boldsymbol{\mathcal{B}}}_m^{\prime \prime}(\mathcal{X}) {\boldsymbol{\mathcal{B}}}_{m^\prime}^{\prime \prime}(\mathcal{X}) d \mathcal{X}$. Coefficient estimation is analogous with a ridge regression, resulting the estimates $\hat{\boldsymbol{b}}=\left({ \boldsymbol{\mathcal{B}} }^{\top} { \boldsymbol{\mathcal{B}} }+\lambda_{\mathcal{B}} \boldsymbol{\Lambda} \right)^{-1} { \boldsymbol{\mathcal{B}} }^{\top} \boldsymbol{S}$ with $\lambda_{\mathcal{B}} = \sigma^2_{\mathcal{B}} q_{\mathcal{B}}$ and $\widehat{\boldsymbol{S}} = \boldsymbol{\mathcal { B }} \widehat{\boldsymbol{b}}$. Spectral analysis of the penalty reveals that $\text{rank}(\boldsymbol{\Lambda})=M+2$, meaning that $M+2$ covariates from $\boldsymbol{\mathcal{B}}$ are penalized, resulting in two fixed effects and $M+2$ random effects. The spectral decomposition yields $\boldsymbol{\Lambda}= \mathbf{P D P}^{\top}$, where $\mathbf{D}=\operatorname{diag}\left( 0,0, d_{1}, \ldots, d_{M+2} \right)$, $\mathbf{P}^{\top} \mathbf{P} =\mathbf{I}_{M+4}$ and $\mathbf{P}=\left( \mathbf{X}_{\Lambda}, \mathbf{Z}_{\Lambda} \right)$. We can write the penalized projection matrix of $\widehat{\boldsymbol{\alpha}}$ as
\begin{equation} \label{eq2}
\begin{array}{rl}
{ \boldsymbol{\mathcal{B}} } \left({ \boldsymbol{\mathcal{B}} }^{\top} { \boldsymbol{\mathcal{B}} }+\lambda_{\mathcal{B}} \boldsymbol{\Lambda} \right)^{-1} { \boldsymbol{\mathcal{B}} }^{\top}
     &=
     { \boldsymbol{\mathcal{B}} } \mathbf{P D}^{-1}_* \mathbf{D}_* \mathbf{P}^{\top} \left( { \boldsymbol{\mathcal{B}} }^{\top} { \boldsymbol{\mathcal{B}} } + \lambda_{\mathcal{B}} \boldsymbol{\Lambda} \right)^{-1} \mathbf{P D}_* \mathbf{D}^{-1}_* \mathbf{P}^{\top} { \boldsymbol{\mathcal{B}} }^{\top} \\
     &= 
     { \boldsymbol{\mathcal{B}} } \mathbf{P D}^{-1}_* \left( \mathbf{D}^{-1}_* \mathbf{P}^{\top} \left( { \boldsymbol{\mathcal{B}} }^{\top} { \boldsymbol{\mathcal{B}} } + \lambda_{\mathcal{B}} \boldsymbol{\Lambda} \right) \mathbf{P D}^{-1}_* \right)^{-1} \mathbf{D}^{-1}_* \mathbf{P}^{\top} { \boldsymbol{\mathcal{B}} }^{\top} \\
     &= 
     \mathbf{C} \left( \mathbf{C}^\top \mathbf{C} + \lambda_{\mathcal{B}} \operatorname{diag}\left(0,0, \mathbf{I}_{M+2} \right) \right)^{-1} \mathbf{C}^\top
\end{array}
\end{equation}
where $\mathbf{D}_*=\operatorname{diag}\left(1,1, \sqrt{d_{1}}, \ldots, \sqrt{d_{M+2}}\right)$, $\mathbf{C} = { \boldsymbol{\mathcal{B}} } \mathbf{P D}^{-1}_* = (\mathbf{X}_{\mathcal{B}}, \mathbf{Z}_{\mathcal{B}})$, $\mathbf{X}_{\mathcal{B}}={ \boldsymbol{\mathcal{B}} } \mathbf{X}_{\Lambda}$, and $\mathbf{Z}_{\mathcal{B}}= { \boldsymbol{\mathcal{B}} } \mathbf{Z}_{\Lambda} \operatorname{diag}\left(d_{1}^{-1 / 2}, \ldots, d_{M+2}^{-1 / 2}\right)$. 

Equation \eqref{eq2}, follows the BLUP form of a mixed effect model with random effects on covariates $\mathbf{Z}_{\mathcal{B}}$ \citep{robinson1991blup,speed1991blup}. In addition, $\mathbf{X}_{\mathcal{B}} \in \operatorname{Span}\left( \left[ \mathbf{1}, \boldsymbol{\mathcal{X}} \right] \right)$, allowing us to substitute $\mathbf{X}_{\mathcal{B}}$ with the original design matrix: $\left[ \mathbf{1}, \boldsymbol{\mathcal{X}} \right]_i = (1,\mathcal{X}_i)$ of an intercept and continuous predictor $\mathcal{X}$. Here, $\mathbf{Z}_{\mathcal{B}}$ is a Demmler-Reinsch (DR) matrix corresponding to the random effects \citep{demmler1975oscillation}. Alternatively, we can replace $\mathbf{D}_*$ with  $\mathbf{D}_{\dagger}=\mathbf{D}^{-1}_{\mathcal{X}} \oplus \operatorname{diag}\left(\sqrt{d_{1}}, \ldots, \sqrt{d_{M+2}}\right)$, where $\mathbf{D}_{\mathcal{X}}$ is a $2\times2$ matrix with column vectors as regression coefficients that map from $\mathbf{X}_{\mathcal{B}}$ to $\left[ \mathbf{1}, \boldsymbol{\mathcal{X}} \right]$, $\mathbf{D}_{\mathcal{X}} = \left( \mathbf{X}^\top_{\mathcal{B}} \mathbf{X}_{\mathcal{B}}\right)^{-1} \mathbf{X}^{\top}_{\mathcal{B}} \left[ \mathbf{1}, \boldsymbol{\mathcal{X}} \right] $. Here, $\oplus$ is a direct sum which concatenates matrices into a block diagonal matrix. Equation \eqref{eq1} can be represented as linear mixed effect models, which can be expressed as $\widehat{\boldsymbol{S}} = \boldsymbol{\mathcal{X}}  \widehat{ {\beta} }_{\mathcal{X}} + \mathbf{Z}_{\mathcal{B}} \widehat{ \mathbf{u} }_{\mathcal{B}}$ with $\mathbf{u} _{\mathcal{B}} \sim \mathrm{N}(\textbf{0}, \tau^{-1}_{\mathcal{B}} \mathbf{I}_{M+2})$ and the intercept being assigned to $\alpha(t)$. 

\subsection{Connection with Bayesian GLMMs}
\label{GLMM}
Use our previous derviation, we can write $ \mathbf{m}_i^\top \boldsymbol{\eta} = \mathbf{x}_i^{\top} \boldsymbol{\beta} + \alpha(t_i) + \mathcal{S}(\mathcal{X}_i) = \mathbf{x}_i^\top \boldsymbol{\beta} + \mathbf{z}_{\alpha,i}^\top \mathbf{u}_{\alpha} + \mathbf{z}_{\mathcal{B},i}^\top \mathbf{u}_{\mathcal{B}} $ in regression matrix form. We can also write the analogous penalized negative log likelihood for GLMMs as 
\begin{equation} \label{pllglm}
- \log L(\boldsymbol{\eta} \mid \mathbf{y}, \mathbf{M} ) + \tau_{\alpha} \| \mathbf{u}_{\alpha} \|^2_2 + \tau_{\mathcal{B}} \| \mathbf{u}_{\mathcal{B}} \|^2_2
\end{equation} 
such that $\mathbf{u}_\alpha \geq \mathbf{0}$. Uniformly ergodic Gibbs samplers have been proposed for Bayesian mixed logistic regression \citep{polson2013bayesian,choi2013polya,wang2018analysis,rao2021block}. We derive a slice sampler that ensures monotonicity of $\alpha(t)$ and can be applied to the general class of Bayesian GLMMs and GAMs. We prove CLT properties for MCMC estimators which allows us to simultaneously ensure monotonicity when estimating $\alpha(t)$ and construct joint bands on functions $\alpha(t)$ and $\mathcal{S}(\mathcal{X})$.

\subsubsection{Truncated gamma and truncated normal distributions}

A Bayesian analog of the mixed effect model are priors $\mathbf{u} \sim \mathrm{N}(\textbf{0}, \tau^{-1} \mathbf{I}_{M+2})$, and $\tau \sim \mathrm{T G} \left(a_0, b_0, \tau_{0}\right)$ where $\tau$ follows a truncated gamma distribution, $\pi \left(\tau \mid a_{0}, b_{0}, \tau_{0}\right)=c_1 \left(\tau_{0}, a_{0}, b_{0}\right)^{-1} \tau^{a_{0}-1} \exp \left(-b_{0} \tau \right) \mathbb{I}\left(\tau \geq \tau_{0}\right)$ where $c_1 \left(\tau_{0}, a_{0}, b_{0}\right)=\int_{\tau_{0}}^{\infty} \tau^{a_{0}-1} \exp \left(-b_{0} \tau\right) d \tau$. In practice $\mathrm{Pr}(\tau_j \leq \tau_0)$ is negligibly small and we set $\tau_0=1000^{-1}$ for our analysis. Bayesian analysis uses the data through the likelihood, to update the prior information; we modify the prior to reflect the monotonicity constraint by means of a truncated normal distribution \citep{li2015efficient}: $\mathbf{u} \sim \mathrm{TN}(\boldsymbol{\mu}, \boldsymbol{\Sigma}, \mathbf{R}, \mathbf{c}, \mathbf{d})$,
$$
\pi (\mathbf{u} \mid \boldsymbol{\mu}, \boldsymbol{\Sigma}, \mathbf{R}, \mathbf{c}, \mathbf{d} )
=
\frac{\exp \left\{-\frac{1}{2}(\mathbf{u}-\boldsymbol{\mu})^{\top} \boldsymbol{\Sigma}^{-1}(\mathbf{u}-\boldsymbol{\mu})\right\}}
{c_2 (\boldsymbol{\mu}, \boldsymbol{\Sigma}, \mathbf{R}, \mathbf{c}, \mathbf{d})} 
\mathbb{I}(\mathbf{c} \leq \mathbf{R} \mathbf{u} \leq \mathbf{d}) 
$$
where $c_2 (\boldsymbol{\mu}, \boldsymbol{\Sigma}, \mathbf{R}, \mathbf{c}, \mathbf{d}) =  \oint_{\mathbf{c} \leq \mathbf{R} \mathbf{u} \leq \mathbf{d}} \exp \left\{-\frac{1}{2}(\mathbf{u}-\boldsymbol{\mu})^{\top} \boldsymbol{\Sigma}^{-1}(\mathbf{u}-\boldsymbol{\mu})\right\} d \mathbf{u}$ and $\mathbf{R}$ is a rotation matrix. Note that half-normal distribution prior of $\mathbf{u}_{\alpha} \sim \mathrm{TN}(\mathbf{0}, \tau^{-1}_\alpha \mathbf{I}_{M+2}, \mathbf{I}_{M+2}, \mathbf{0}, \boldsymbol{\infty} )$ preserves the conjugacy of $\tau_\alpha$
$$
\pi (\mathbf{u} \mid \mathbf{0}, \tau^{-1}_\alpha \mathbf{I}_{M+2}, \mathbf{I}_{M+2}, \mathbf{0}, \boldsymbol{\infty} )
=
\left( {2 \tau_\alpha}/{\pi} \right)^{(M+2)/2}
{\exp \left( -\frac{ \tau_\alpha }{2} \mathbf{u}^{\top} \mathbf{u}\right) }
\mathbb{I}(\mathbf{0} \leq \mathbf{u} ) .
$$ 
Because the distribution is zero centered and the covariance matrix is isotropic, each marginal half-normal kernel has half the volume of the normal kernel. A normalization factor of 2 is multiplied to each marginal normal PDF in order to obtain the PDF of the half-normal distribution.

\subsubsection{Block slice sampler}
\label{slice}
The likelihood of exponential family GLMs is given by
$$
L(\boldsymbol{\eta} \mid \mathbf{y}, \mathbf{M} ) = \exp \left( \mathbf{y}^{\top} \mathbf{M} \boldsymbol{\eta} \right) \exp \left( -\sum_{i=1}^{N} \xi \left(\mathbf{m}_{i}^{\top} \boldsymbol{\eta}\right) \right)
$$
where $\xi (v)=e^{v}$ for Poisson regression, $\xi (v)=\log \left(1+e^{v}\right)$ for logistic regression, $\xi (v)=v^{2} / 2$ for linear model, etc. \citep{ghosal2022bayesian}. For the exponential proportion hazards model we have $\xi (v,t)=\exp(\log(t)+v)$, $\exp \left( \mathbf{y}^{\top} \mathbf{X} \boldsymbol{\beta} \right) \exp \left( -\sum_{i=1}^{N} \exp \left( \log(t_i) + \mathbf{x}_{i}^{\top} \boldsymbol{\beta}\right) \right)$. Note that, $\xi (v) \geq 0$ and $\xi (v)$ is convex.

The posterior kernel with normal priors for $\boldsymbol{\eta} \sim \mathrm{TN}( \mathbf{b}, \mathbf{A}(\boldsymbol{\tau})^{-1}, \mathbf{R}_\alpha, \mathbf{c}_\alpha, \boldsymbol{\infty} )$ is
$$
\pi(\boldsymbol{\eta} \mid \mathbf{y}) \propto L(\boldsymbol{\eta} \mid \mathbf{y}, \mathbf{M} )  \pi (\boldsymbol{\eta} \mid \mathbf{b}, \mathbf{A}(\boldsymbol{\tau})^{-1}, \mathbf{R}_\alpha, \mathbf{c}_\alpha, \boldsymbol{\infty} )
$$
where $\mathbf{A}(\boldsymbol{\tau})^{-1}$ is the covariance matrix and $\boldsymbol{\tau}=\{ \tau_\alpha, \tau_\mathcal{B} \}$. If we order $\mathbf{M}=[\mathbf{X}, \mathbf{Z}_\alpha, \mathbf{Z}_\mathcal{B} ]$, $\boldsymbol{\eta}=[\boldsymbol{\beta}^{\top}, \mathbf{u}^{\top}_\alpha, \mathbf{u}^{\top}_\mathcal{B}]^{\top}$, then $\mathbf{A}(\boldsymbol{\tau}) = \boldsymbol{\Sigma}^{-1} \oplus \tau_\alpha \mathbf{I}_{M+2} \oplus \tau_{\mathcal{B}}\mathbf{I}_{M+2}$ is the prior precision. The prior mean is $\mathbf{b} = [\boldsymbol{\mu}^\top, \mathbf{0}_{M+2}^\top, \mathbf{0}_{M+2}^\top ]^{\top}$. We denote the rotation $\mathbf{R}_\alpha =  \mathbf{I}_{p} \oplus \mathbf{I}_{M+2} \oplus \mathbf{I}_{M+2}$ and lower bound as $\mathbf{c}_\alpha = \left[ -\boldsymbol{\infty}^\top_{p}, \mathbf{0}^\top_{M+2}, -\boldsymbol{\infty}^\top_{M+2} \right]^\top$ inorder to ensure $\mathbf{u}_\alpha \geq \mathbf{0}$. Any other linear inequality constraints on the fixed effect coefficients $\boldsymbol{\beta}$ can be concatenated into $\{ \mathbf{R}_\alpha, \mathbf{c}_\alpha \}$, e.g. simplex constraint $\mathbf{1}^\top \boldsymbol{\beta} \geq 1$, $-\mathbf{1}^\top \boldsymbol{\beta} \geq -1$, $\boldsymbol{\beta} \geq \mathbf{0}$. 

We introduce uniformly distributed latent auxiliary variables $\omega_i \sim \mathrm{U}(0,1)$ and inequality constraints on $ \mathbf{u}_\alpha $ to obtain joint posterior 
\begin{equation} \label{SS_joint}
\begin{array}{rl}
\pi(\boldsymbol{\eta}, \boldsymbol{\omega} \mid \mathbf{y}) \propto&
L(\boldsymbol{\eta} \mid \mathbf{y}, \mathbf{M} )  \pi (\boldsymbol{\eta} \mid \mathbf{b}, \mathbf{A}(\boldsymbol{\tau})^{-1}, \mathbf{R}_\alpha, \mathbf{c}_\alpha, \boldsymbol{\infty} ) 
\pi(\boldsymbol{\omega} \mid \boldsymbol{\eta}) \\
\propto &
\exp \left\{\mathbf{y}^{\top} \mathbf{M} \boldsymbol{\eta}-
\frac{1}{2}\left(\boldsymbol{\eta}- \mathbf{b} \right)^{\top} 
\mathbf{A}(\boldsymbol{\tau})
\left(\boldsymbol{\eta}- \mathbf{b} \right)\right\} \\
&\times 
\mathbb{I} ( \mathbf{c}_\alpha \leq \mathbf{R}_\alpha \boldsymbol{\eta} )
\prod_{i=1}^{N} \mathbb{I} \left( \omega_i \leq \exp \left(- \xi \left( \mathbf{m}^\top_i \boldsymbol{\eta} \right) \right) \right) .
\end{array}
\end{equation}
When the joint distribution \eqref{SS_joint} is integrated with respect to $\omega_i$, we obtain the marginal distribution $\pi(\boldsymbol{\eta} \mid \mathbf{y})$ and see that we have $\omega_i|\boldsymbol{\eta} \sim \mathrm{U} \left( 0, \exp \left(- \xi \left( \mathbf{m}^\top_i \boldsymbol{\eta} \right) \right) \right)$ where truncated uniform sampling gave rise to the name slice sampler \citep{mira2002efficiency,damlen1999gibbs,neal2003slice}. In our two variable example for $\{ \boldsymbol{\eta}, \boldsymbol{\omega} \}$, we can show that the mean and conditional distribution of $\boldsymbol{\eta}$ is another truncated normal with the kernel of $\boldsymbol{\eta} | \boldsymbol{\omega}, \boldsymbol{\tau} \sim \mathrm{N} \left( \mathbf{b} + \mathbf{A}(\boldsymbol{\tau})^{-1} \mathbf{M}^{\top} \mathbf{y}, \mathbf{A}(\boldsymbol{\tau})^{-1} \right)$ such that $\prod_{i=1}^{N} \mathbb{I} \left( \omega_i \leq \exp \left(- \xi \left( \mathbf{m}^\top_i \boldsymbol{\eta} \right) \right) \right)$ and $\mathbb{I} ( \mathbf{c}_\alpha \leq \mathbf{R}_\alpha \boldsymbol{\eta} )$ are satisfied.

Incorporating our random effect structure using a gamma distribution truncated below at $\tau_0$, $\tau_j \sim \mathrm{T G} \left(a_0, b_0, \tau_{0} \right)$, our joint distribution becomes
$$
\begin{array}{rl}
\pi(\boldsymbol{\eta}, \boldsymbol{\omega}, \boldsymbol{\tau} \mid \mathbf{y}) \propto&
L(\boldsymbol{\eta} \mid \mathbf{y}, \mathbf{M} )  \pi (\boldsymbol{\eta} \mid \mathbf{b}, \mathbf{A}(\boldsymbol{\tau})^{-1}, \mathbf{R}_\alpha, \mathbf{c}_\alpha, \boldsymbol{\infty} ) 
\pi(\boldsymbol{\omega} \mid \boldsymbol{\eta}) 
\pi \left( \boldsymbol{\tau} \mid a_{0}, b_{0}, \tau_{0}\right) \\
\propto & 
\exp \left\{\mathbf{y}^{\top} \mathbf{M} \boldsymbol{\eta}-
\frac{1}{2}\left(\boldsymbol{\eta}- \mathbf{b} \right)^{\top} 
\mathbf{A}(\boldsymbol{\tau})
\left(\boldsymbol{\eta}- \mathbf{b} \right)\right\} \\
&\times \prod_{j \in \{\alpha, \mathcal{B}\}} \tau_j^{a_{0}+M/2} e^{-b_{0} \tau_j} \mathbb{I}\left(\tau_j \geq \tau_{0}\right) \\
&\times 
\mathbb{I} ( \mathbf{c}_\alpha \leq \mathbf{R}_\alpha \boldsymbol{\eta} )
\prod_{i=1}^{N} \mathbb{I} \left( \omega_i \leq \exp \left(- \xi \left( \mathbf{m}^\top_i \boldsymbol{\eta} \right) \right) \right) .
\end{array}
$$
The conditional distributions $\tau_j \mid \boldsymbol{\eta} \sim \mathrm{T G} \left(a_{0}+({M+2})/{2}, b_{0}+{ \mathbf{u}_j^{\top} \mathbf{u}_j}/{2}, \tau_{0}\right)$ can be derived from the joint distribution. Prior for $\tau_\alpha$ induces shrinkage on $\alpha(t)$ and prior $\tau_{\mathcal{B}}$ induces smoothness on $\mathcal{S}(\mathcal{X})$ based on the second derivative. Note the joint distribution of $\{ \boldsymbol{\eta}, \boldsymbol{\tau} \}$ is given by 
$$
\pi(\boldsymbol{\eta}, \boldsymbol{\tau} \mid \mathbf{y}) \propto
L(\boldsymbol{\eta} \mid \mathbf{y}, \mathbf{M} ) \pi (\boldsymbol{\eta} \mid \mathbf{b}, \mathbf{A}(\boldsymbol{\tau})^{-1}, \mathbf{R}_\alpha, \mathbf{c}_\alpha, \boldsymbol{\infty} ) \pi \left( \boldsymbol{\tau} \mid a_{0}, b_{0}, \tau_{0}\right) .
$$

Here $\xi(v)$ is convex and lower bounded at 0 which allows us to obtain linear inequality constraints for $\boldsymbol{\eta}$ given $\omega_i \in (0,1)$
\begin{align*}
\omega_i &\leq \exp( - \xi \left( \mathbf{m}^\top_i \boldsymbol{\eta} \right) ) = \exp( - \log (1 + \exp(\mathbf{m}^\top_i \boldsymbol{\eta}) ) ) < 1 \\
\omega^{-1}_i &\geq 1 + \exp \left( \mathbf{m}^\top_i \boldsymbol{\eta} \right) \\
\log(\omega_i^{-1} - 1) &\geq \mathbf{m}^\top_i \boldsymbol{\eta} .
\end{align*}
We can concatenate $\mathbf{u}_\alpha \geq \mathbf{0}$ and $-\mathbf{m}^\top_i \boldsymbol{\eta} \geq -\log(\omega_i^{-1} - 1)$ for all $i$ as a rotation matrix inequality, $\mathbf{R}_\omega \boldsymbol{\eta} \geq \mathbf{c}_\omega$. Here we stack $-\mathbf{M}$ on top of $\mathbf{R}_\alpha$ to get $\mathbf{R}_\omega$. We define vector ${[\mathbf{c}_*]}_i = -\log(\omega_i^{-1} - 1)$ and stack it on top of $\mathbf{c}_\alpha$ to get $\mathbf{c}_\omega$. Note that $\mathbf{c}_\omega$ is a function of $\boldsymbol{\omega}$. Our block Gibbs sampler is given as 
\begin{equation} \label{S_sampler}
\begin{array}{c}
\boldsymbol{\eta} \mid \boldsymbol{\omega}, \boldsymbol{\tau} \sim \mathrm{TN} \left( \mathbf{A}(\boldsymbol{\tau})^{-1}\boldsymbol{\mu}_*, \mathbf{A}(\boldsymbol{\tau})^{-1}, \mathbf{R}_\omega, \mathbf{c}_\omega, \boldsymbol{\infty} \right) \\
\omega_i \mid \boldsymbol{\eta} \sim \mathrm{U} \left( 0, \exp( - \xi (\mathbf{m}^\top_i \boldsymbol{\eta}) ) \right) \\
\tau_j \mid \boldsymbol{\eta} \sim \mathrm{T G} \left(a_{0}+\frac{M+2}{2}, b_{0}+\frac{ \mathbf{u}^{\top}_j \mathbf{u}_j}{2}, \tau_{0}\right)
\end{array}
\end{equation} 
where $\xi(v) = \log(1+e^v)$ and $\boldsymbol{\mu}_* = \mathbf{M}^{\top} \mathbf{y} + \mathbf{A}(\boldsymbol{\tau}) \mathbf{b} = \mathbf{M}^{\top} \mathbf{y} + \left[ \left[ \boldsymbol{\Sigma}^{-1} \boldsymbol{\mu} \right]^{\top}, \mathbf{0}^\top_{2M+4} \right]^{\top}$. Note that \eqref{S_sampler} is a general Gibbs sampler for linear inequality constrained Bayesian GLMMs and GAMs which can be customized based on $\xi(v)$ and shape determining basis system \citep{meyer2008inference,ghosal2023shape}.

\section{Uniform ergodicity and Markov chain central limit theorem}
We establish uniform ergodicity and that posterior samples are square integrable i.e., the second moment exist which guarantees central limit theorem (CLT) results for posterior averages and consistent estimators of the associated asymptotic variance \citep{jones2004markov}. This Markov chain CLT result is a special case of martingale CLT \citep{kurtz1981central,meyn2012markov}. Our posterior samples for our functions $\alpha(t)$ and $\mathcal{S}(\mathcal{X})$ are matrix multiplication of the posterior coefficient samples, $\mathcal{Z}_\alpha \mathbf{u}_\alpha$ and $\mathcal{Z}_\mathcal{B} \mathbf{u}_\mathcal{B}$ and have the same CLT properties. Here $\mathcal{Z}_\alpha$ and $\mathcal{Z}_\mathcal{B}$ are matrix representation of the continuous I-spline and Demmler-Reinsch bases.

First, we show uniform ergodicity of $\boldsymbol{\eta}$ from Gibbs sampler \eqref{S_sampler}, taking advantage of the truncated gamma \citep{wang2018analysis}. A key feature of this strategy is that by truncating the gamma distribution at a small $\tau_0 = \epsilon$ results in useful inequalities related to ergodicity and in practice $\mathrm{Pr}(\tau_j \leq \epsilon)$ is negligibly small. We denote the $\boldsymbol{\eta}$-marginal Markov chain as $\Psi \equiv \{ \boldsymbol{\eta}(n) \}^\infty_{n=0}$ and Markov transition density (Mtd) of $\Psi$ as 
$$
k\left( \boldsymbol{\eta} \mid \boldsymbol{\eta}^{\prime}\right)
=
\int_{\mathbb{R}_{+}} \int_{\mathbb{H}^{N}} \pi( \boldsymbol{\eta} \mid \boldsymbol{\omega}, \boldsymbol{\tau}, \mathbf{y}) \pi\left(\boldsymbol{\omega}, \boldsymbol{\tau} \mid \boldsymbol{\eta}^{\prime}, \mathbf{y}\right) d \boldsymbol{\omega} d {\tau}
$$
where $\boldsymbol{\eta}^{\prime}$ is the current state and $\boldsymbol{\eta}$ is the next state, with $\mathbb{H}^{N}=[0,1]^N$ is a hypercube. We show the Mtd of satisfies the following minorization condition: $k\left( \boldsymbol{\eta} \mid \boldsymbol{\eta}^{\prime}\right) \geq \delta h(\boldsymbol{\eta})$, where there exist a $\delta > 0$ and density function $h$, to prove uniform ergodicity \citep{roberts2004general}. Uniform ergodicity is  defined as bounded and geometrically decreasing bounds for total variation distance to the stationary distribution in number of Markov transitions $n$, $\left\|K^{n}( \boldsymbol{\eta}, \cdot)-\Pi(\cdot \mid \mathbf{y})\right\|:=\sup _{A \in \mathscr{B}}\left|K^{n}( \boldsymbol{\eta}, A)-\Pi(A \mid \mathbf{y})\right| \leq V r^{n}$.
Here $\mathscr{B}$ denotes the Borel $\sigma$-algebra of $\mathbb{R}^{p+2M+4}$, $K(\cdot, \cdot)$ be the Markov transition function for the Mtd $k(\cdot, \cdot)$ 
$$
K\left( \boldsymbol{\eta}^{\prime}, A\right)=\operatorname{Pr}\left( \boldsymbol{\eta}^{(j+1)} \in A \mid \boldsymbol{\eta}^{(j)}=\boldsymbol{\eta}^{\prime}\right)=\int_{A} k\left( \boldsymbol{\eta} \mid \boldsymbol{\eta}^{\prime}\right) d \boldsymbol{\eta},
$$
and $K^{n}\left( \boldsymbol{\eta}^{\prime}, A\right)=\operatorname{Pr}\left( \boldsymbol{\eta}^{(n+j)} \in A \mid \boldsymbol{\eta}^{(j)}=\boldsymbol{\eta}^{\prime}\right)$
We denote $\Pi(\cdot \mid \mathbf{y})$ as the probability measure with density $\pi(\boldsymbol{\eta} \mid \mathbf{y})$, $V$ is bounded above and $r\in(0,1)$.

\begin{theorem} \label{thm1}
Assume that $a_0 > 0$, $b_0 > 0$, $\xi ( \mathbf{m}_i^{\top} \boldsymbol{\eta} ) \geq 0$ and $\xi(v)$ is convex, then the Markov chain $\Psi$ of \eqref{S_sampler} for constrained Bayesian GLMMs and GAMs is uniformly ergodic.
\end{theorem}

\begin{theorem} \label{thm2}
For any fixed $\mathbf{t} \in \mathbb{R}^{p+2M+4}$, $\int_{\mathbb{R}^{p+2M+4}} e^{\boldsymbol{\eta}^{\top} \mathbf{t}} \pi(\boldsymbol{\eta} \mid \mathbf{y}) d \boldsymbol{\eta}<\infty$. Hence, the moment generating function of the posterior distribution exists.
\end{theorem}

The prior precision is lower bounded, making the variances of mixed effect components upper bounded and contained in a hypercube. By transiting our parameters of interest through a flexible auxiliary variable space, bounded in a finite volume hypercube, we obtain desirable integral properties for our Markov chain. We leave the proofs to the Proofs section. Note that the proof requires $a_0 + (M+2)/2  \geq 1$ which is implied by the knot selection and construction of the spline bases. This condition is necessary when adapting our slice sampler to a general mixed model setting.

\subsection{Posterior inference: Joint bands, SimBaS and GBPV}

Our CLT properties extend to continuous functional predictors, allowing us to construct joint bands while accounting for a Bayesian false discovery rate. Using $\alpha(t)$ as an example, the Simultaneous Band Scores (SimBaS) and joint bands found in \cite{meyer2015bayesian} and \cite{ruppert2003semiparametric} are a direct corollary of our CLT results. Suppose $\alpha^{(r)}(t) = \beta^{(r)}_{0} + \mathcal{Z}_\alpha \mathbf{u}_\alpha^{(r)}$ for $r\in \{1,\dots,R\}$ is a sample from the posterior, then intervals for $\alpha(t)$ are given as 
$$
I_{u}(t)=\hat{\alpha}(t) \pm q_{(1-u)}[\widehat{\operatorname{St.Dev}}\{\hat{\alpha}(t)\}]
$$
where the variable $q_{(1-u)}$ is the $(1-u)$ quantile taken over $R$ of the quantity
$$
Z^{(r)}= \max_{ t \in \mathcal{T} } \left|\frac{\alpha^{(r)}(t)-\hat{\alpha}(t)}{\widehat{\operatorname{St.Dev}}\{\hat{\alpha}(t)\}}\right|
$$
and multiplicity adjusted probability score at values of $t$ are given as
$$
P_{\operatorname{SimBaS}}(t)=\frac{1}{R} \sum_{r=1}^{R} \mathbb{I} \left\{\left|\frac{\hat{\alpha}(t)}{\widehat{\operatorname{St} \cdot \operatorname{Dev}}\{\hat{\alpha}(t)\}}\right| \leq Z^{(r)}\right\} .
$$
For each $t$, $\mathrm{P}_{\operatorname{SimBaS}}(t)$ can be used as local probability scores that have multiple testing adjusted global properties. For example, we can flag domain $\left\{t: \mathrm{P}_{\operatorname{SimBaS}}(t)<u\right\}$ as significant. From these we can compute $P_{\text{Bayes}}=\min_t\left\{P_{\operatorname{SimBaS}}(t)\right\}$, which denote global Bayesian p-values (GBPV) such that we reject the global hypothesis that $\alpha(t) \equiv 0$ whenever $P_{\text{Bayes}}<u$.

\section{Illustrative examples}
\begin{figure}
\centering
\includegraphics[width=6in]{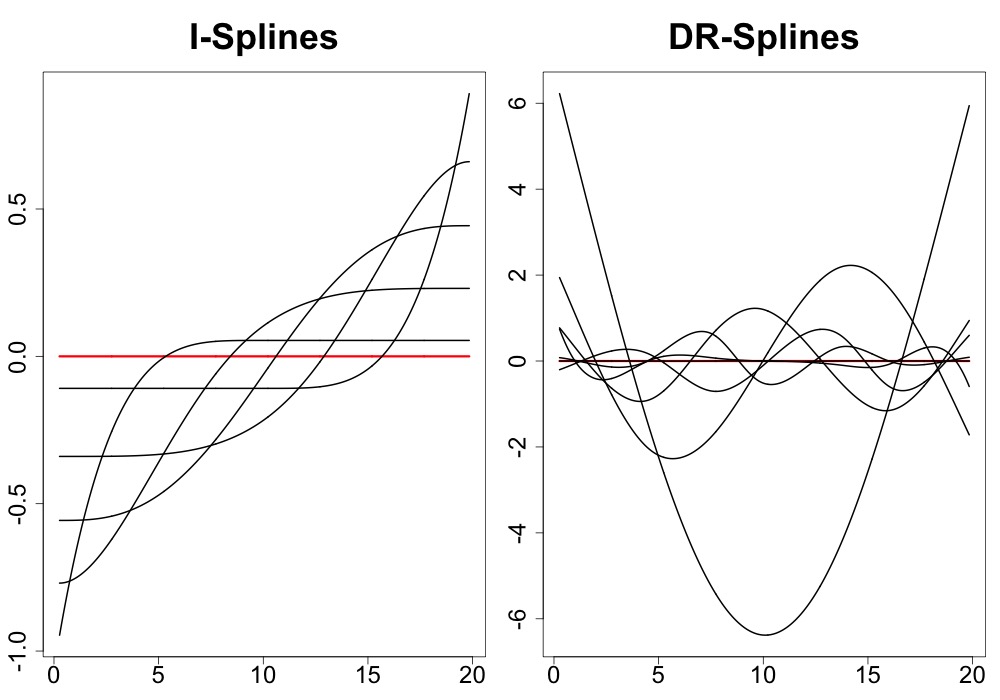}
\caption{Examples of I-Spline and DR-Spline bases for $\alpha(t)$ and $\mathcal{S}(\mathcal{X})$. }
\label{fig:1}
\end{figure}

We use the \texttt{R} package \texttt{cascsim} \cite{cascsim} to sample from truncated gamma and \texttt{tmvtnsim} \cite{tmvtnsim} efficiently sample from truncated normal. The package \texttt{tmvtnsim} is a C++ implementation of algorithms found in \cite{li2015efficient}. In addition, we follow the construction of I-splines found in \cite{meyer2008inference} and are normalized such that $\mathbf{Z}_{\alpha}^\top \mathbf{y} = \mathbf{0}$ and are plotted in Figure \ref{fig:1}.

\begin{figure}
\centering
\includegraphics[width=6in]{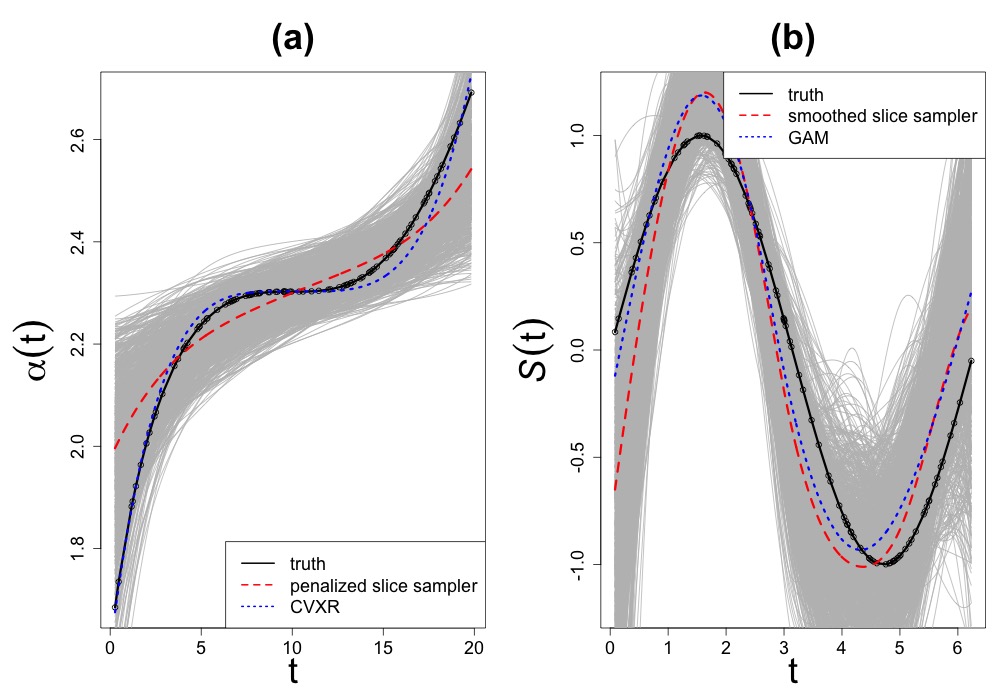}
\caption{Poisson models fitted with I-splines and DR-bases: a) unpenalized vs penalized I-splines using the convex optimization and slice sampler; b) second derivative penalty with GAMs and slice sampler. Slice sampler shrinks the monotonic effect towards the intercept. Smoothed GAM and slice sampler have similar fits. Posterior samples are plotted with gray lines.}
\label{fig:2}
\end{figure}

We simulate and fit two Poisson models for our illustrative examples: a) monotonic regression with log mean, $\log(\mu(t)) = \alpha(t) = \log( 0.005(t-10)^3+10 )$; b) nonparametric regression with log mean, $\log(\mu(t)) = \mathcal{S}(t) = \sin(t)$. For model a), we maximize the unpenalized constrained likelihood of \eqref{pllglm} using \texttt{R} package \texttt{CVXR} \citep{fu2020cvxr} and penalized version of the likelihood using the slice sampler and observe that the slice sampler shrinks the monotonic covariate effect towards the intercept. Our hierarchical modeling of I-spline coefficients adaptively shrinks the monotonic nonparametric effect. For our second illustrative example, model b), we compare the slice sampler to the widely used GAMs implemented in \texttt{mgcv} \citep{wood2022mgcv}. Both \texttt{mgcv} and the slice sampler uses a second derivative smoothness penalty and we observed similar model fits for both methods. We plot our results in Figure \ref{fig:2} and used vague priors for our analyses. 

\section{Real data analysis}
\begin{figure}
\centering
\includegraphics[width=6in]{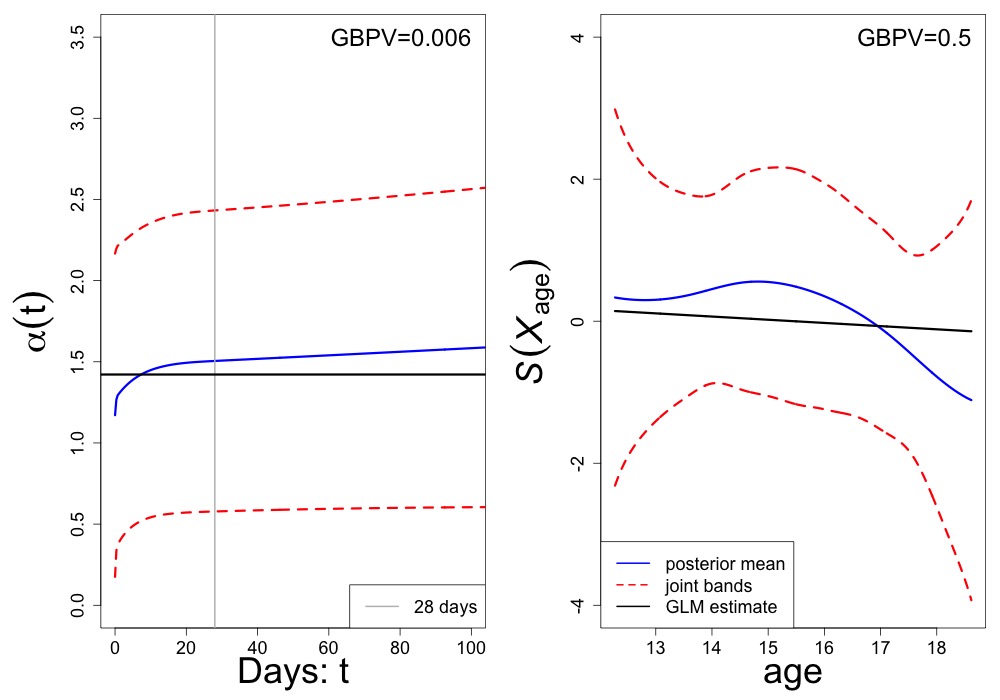}
\caption{Global Bayesian p-values (GBPVs), posterior mean, 0.95 joint bands for monotonic time effect and nonparametric age effect and GLM estimates. Logistic regression GLM was fitted without nonparametric effects for comparison. Steep increases in the probability of recovery as a function of time occurs before 28 days. Relatively constant odds of recovery are observed for age 12-16 with odds of recovery decreasing after age 16.}
\label{fig:3}
\end{figure}

We fit a current status proportional odds model for concussion recovery data from a study conducted at the Children's Hospital of Pennsylvania. A sample of $N=74$ participants from a large prospective observational cohort study assessing diagnostic measures of concussion were used in our analysis. In this example, $Y$ is the whether a subject recovered by the last clinical visit and $T$ is the time from concussion to last clinic visit. Recovery, $Y$ is defined as a SCAT5 score less than 5 at the time of last clinic visit \citep{echemendia2017sport}. We include linear effects for King-Devick (KD) test completion time \citep{galetta2015adding} and pupil eye pain binary variable, both recorded at the last clinic visit. In addition, we include age as a nonparametric effect. Age and KD completion time were mean centered and we used knots based on quantiles of the raw data distribution to generate $M+2=7$ splines bases for mixed effect components. We fit $\operatorname{logit} (F(t_i \mid \mathbf{x}_i ) )=\alpha(t_i) + \mathcal{S}(\mathcal{X}_{\text{age}_i}) + {x}_{\text{pain}} \beta_{\text{pain}} + {x}_{\text{KD}} \beta_{\text{KD}}$ using our slice sampler with results for nonparametric effects plotted in Figure \ref{fig:3}. We expect a positive intercept and $\alpha(t)>0$ due to 73\% of our cohort being recovered cases. However, we observed steep increases in the probability of recovery as a function of time, before 28 days. In addition, we observed that recovery odds, $\mathcal{S}(\mathcal{X}_{\text{age}_i})$ remains relatively constant until age 16 then decreases with age. These results are aligned with known factors associated with concussion recovery with nonparametric regression elucidating the shape of these trends \citep{desai2019factors}. In addition, the posterior mean and 0.95 credible interval for pupil eye pain effect is $-3.267(-5.632, -1.214)$ and $-0.043(-0.111, 0.019)$ for KD completion time effect. Our model suggest that pupil eye pain is highly correlated with having a concussion and pupil eye pain is already widely used as a concussion indicator. Longer time spent completing the KD test is associated with a higher probability of concussion; however, this relationship remains marginal in our analysis.

\section{Discussion} 
Our Markov chain CLT results establishes rigorous inference with finite sample data and settings with shape and linear inequality constraints. Our MCMC procedure allows for simultaneous estimation and inference unlike traditional constrained convex optimization methods. In addition, we derived second derivative smoothness penalties through random effect parameterization which allows GAMs to be fitted with our slice sampler. The hierarchical modeling of random effects also allow us to adaptively induce shrinkage on nonparametric monotone effects. Our mixed effect slice sampler can be applied to canonical exponential family models which includes GLMM, GAM and Exponential PH models.

One limitation of our approach is sampling from high dimensional truncated normal distribution can be slow. Possible resolutions for high dimensional settings are parallel MCMC methods \citep{xing2014asymp}. In addition, we hope that increase use of slice samplers can lead to new research in truncated normal sampling. Our Gibbs sampler and proof techniques can be adapted to related settings such as Bayesian variable selection and is a promising direction for future work. 

\section*{Proofs}
\begin{proof}Proof of Theorem \ref{thm1}.
Here we rely on the properties of slice samplers and useful inequalities due to the truncated gamma prior. Note that $\mathbf{A}\left(\tau_{0}\right) = \boldsymbol{\Sigma}^{-1} \oplus \tau_0 \mathbf{I}_{M+2} \oplus \tau_0\mathbf{I}_{M+2}$, $\mathbf{A}\left(\boldsymbol{\tau}\right) \succeq \mathbf{A}\left(\tau_{0}\right)$, $\exp \left( -\xi \left( \mathbf{m}^\top_i \boldsymbol{\eta} \right) \right) \leq 1$, $g(\boldsymbol{\eta} ) \mathbb{I}( \mathbf{c}_\alpha \leq \mathbf{R}_\alpha \boldsymbol{\eta} ) \leq g(\boldsymbol{\eta} )$ and $\oint_{ \mathbf{R}_\alpha \boldsymbol{\eta} \geq \mathbf{c}_\alpha } g( \boldsymbol{\eta} ) d \boldsymbol{\eta} = \int_{\mathbb{R}^{p+2M+4}} g( \boldsymbol{\eta} ) \mathbb{I}( \mathbf{c}_\alpha \leq \mathbf{R}_\alpha \boldsymbol{\eta} ) d \boldsymbol{\eta} \leq \int_{\mathbb{R}^{p+2M+4}} g(\boldsymbol{\eta} ) d \boldsymbol{\eta}$.

We have
$$
\begin{array}{rl}
\pi( \boldsymbol{\eta} \mid \boldsymbol{\omega}, \boldsymbol{\tau}, \mathbf{y}) \pi\left(\boldsymbol{\omega} \mid \boldsymbol{\eta}^{\prime}, \mathbf{y}\right) 
= &
\exp \left\{-\frac{1}{2}( \boldsymbol{\eta}- \mathbf{A}(\boldsymbol{\tau})^{-1}\boldsymbol{\mu}_* )^{\top} \mathbf{A}(\boldsymbol{\tau})( \boldsymbol{\eta}- \mathbf{A}(\boldsymbol{\tau})^{-1}\boldsymbol{\mu}_* )\right\} \\
& \times 
\frac{\mathbb{I}(\mathbf{c}_\omega \leq \mathbf{R}_\omega \boldsymbol{\eta} )}
{c_2 ( \mathbf{A}(\boldsymbol{\tau})^{-1}\boldsymbol{\mu}_*, \mathbf{A}(\boldsymbol{\tau})^{-1}, \mathbf{R}_\omega, \mathbf{c}_\omega, \infty)} \\
& \times
\prod_{i=1}^{N} \mathbb{I} \left( \omega_i \leq \exp \left(- \xi \left( \mathbf{m}^\top_i \boldsymbol{\eta}^\prime \right) \right) \right) \exp \left( \xi \left( \mathbf{m}^\top_i \boldsymbol{\eta}^\prime \right) \right) \\
= &
\exp \left[-\frac{1}{2} \left( \boldsymbol{\eta}^{\top} \mathbf{A}(\boldsymbol{\tau}) \boldsymbol{\eta} - 2  \boldsymbol{\eta}^{\top} \boldsymbol{\mu}_* \right) - \frac{1}{2} \boldsymbol{\mu}^{\top}_* \mathbf{A}(\boldsymbol{\tau})^{-1} \boldsymbol{\mu}_* \right] \\
& \times 
\frac{ \mathbb{I}(\mathbf{c}_\omega \leq \mathbf{R}_\omega \boldsymbol{\eta} ) }
{c_2 ( \mathbf{A}(\boldsymbol{\tau})^{-1}\boldsymbol{\mu}_*, \mathbf{A}(\boldsymbol{\tau})^{-1}, \mathbf{R}_\omega, \mathbf{c}_\omega, \infty)} \\
& \times 
\prod_{i=1}^{N} \mathbb{I} \left( \omega_i \leq \exp \left(- \xi \left( \mathbf{m}^\top_i \boldsymbol{\eta}^\prime \right) \right) \right) \exp \left( \xi \left( \mathbf{m}^\top_i \boldsymbol{\eta}^\prime \right) \right) \\
= &
\exp \Bigg[ -\frac{1}{2} \left( \boldsymbol{\beta}^{\top} \boldsymbol{\Sigma}^{-1} \boldsymbol{\beta} + \sum_{j \in \{ \alpha, \mathcal{B} \}} \tau_j \mathbf{u}_j^{\top} \mathbf{u}_j - 2  \boldsymbol{\eta}^{\top} \boldsymbol{\mu}_* \right) \\
& - \frac{1}{2} \boldsymbol{\mu}^{\top}_* \mathbf{A}(\boldsymbol{\tau})^{-1} \boldsymbol{\mu}_* \Bigg] \\
& \times 
\frac{ \mathbb{I}(\mathbf{c}_\omega \leq \mathbf{R}_\omega \boldsymbol{\eta} ) }
{c_2 ( \mathbf{A}(\boldsymbol{\tau})^{-1}\boldsymbol{\mu}_*, \mathbf{A}(\boldsymbol{\tau})^{-1}, \mathbf{R}_\omega, \mathbf{c}_\omega, \infty)} \\
& \times 
\prod_{i=1}^{N} \mathbb{I} \left( \omega_i \leq \exp \left(- \xi \left( \mathbf{m}^\top_i \boldsymbol{\eta}^\prime \right) \right) \right) \exp \left( \xi \left( \mathbf{m}^\top_i \boldsymbol{\eta}^\prime \right) \right) \\
\end{array}
$$
where
$$
\begin{array}{rl} 
& c_2 \left(\mathbf{A}(\boldsymbol{\tau})^{-1} \boldsymbol{\mu}_*, \mathbf{A}(\boldsymbol{\tau})^{-1}, \mathbf{R}_\omega, \mathbf{c}_\omega, \infty \right) \\
&=  \oint_{ \mathbf{R}_\omega \boldsymbol{\eta} \geq \mathbf{c}_\omega } \exp 
\left[
-\frac{1}{2}( \boldsymbol{\eta} - \mathbf{A}(\boldsymbol{\tau})^{-1} \boldsymbol{\mu}_* )^{\top} 
 \mathbf{A}(\boldsymbol{\tau})
( \boldsymbol{\eta} - \mathbf{A}(\boldsymbol{\tau})^{-1} \boldsymbol{\mu}_*)
\right] 
d \boldsymbol{\eta} \\
&\leq  \oint_{ \mathbf{R}_\omega \boldsymbol{\eta} \geq \mathbf{c}_\omega } \exp 
\left[
-\frac{1}{2}( \boldsymbol{\eta} - \mathbf{A}(\boldsymbol{\tau})^{-1} \boldsymbol{\mu}_* )^{\top} 
\mathbf{A}(\tau_0) 
( \boldsymbol{\eta} - \mathbf{A}(\boldsymbol{\tau})^{-1} \boldsymbol{\mu}_\omega)
\right] 
d \boldsymbol{\eta} \\
&< \int_{\mathbb{R}^{p+2M+4}} \exp 
\left[
-\frac{1}{2}( \boldsymbol{\eta} - \mathbf{A}(\boldsymbol{\tau})^{-1} \boldsymbol{\mu}_\omega )^{\top} 
\mathbf{A}(\tau_0) 
( \boldsymbol{\eta} - \mathbf{A}(\boldsymbol{\tau})^{-1} \boldsymbol{\mu}_\omega)
\right] 
d \boldsymbol{\eta} \\
&= (2 \pi)^{(p+2M+4)/2} \left| \mathbf{A}(\tau_0) \right|^{-1/2} .
\end{array}
$$
Note that $\mathbb{I}(\mathbf{c}_\omega \leq \mathbf{R}_\omega \boldsymbol{\eta} ) = \mathbb{I}(\mathbf{c}_\alpha \leq \mathbf{R}_\alpha \boldsymbol{\eta} ) \prod_{i=1}^{N} \mathbb{I} \left( \omega_i \leq \exp \left(- \xi \left( \mathbf{m}^\top_i \boldsymbol{\eta} \right) \right) \right)$.
From \cite{mira2002efficiency} (Theorem 7), we have
$$
\begin{array}{rl}
&\displaystyle\int_{\mathbb{H}^{N}}  \pi( \boldsymbol{\eta} \mid \boldsymbol{\omega}, \boldsymbol{\tau}, \mathbf{y}) \pi\left(\boldsymbol{\omega} \mid \boldsymbol{\eta}^{\prime}, \mathbf{y}\right) 
d \boldsymbol{\omega} \\
& \geq
\exp \left[-\frac{1}{2} \left( \boldsymbol{\eta}^{\top} \mathbf{A}(\boldsymbol{\tau}) \boldsymbol{\eta} - 2  \boldsymbol{\eta}^{\top} \boldsymbol{\mu}_* \right) - \frac{1}{2} \boldsymbol{\mu}^{\top}_* \mathbf{A}(\boldsymbol{\tau})^{-1} \boldsymbol{\mu}_* \right] \\ 
& \times 
\frac{ \mathbb{I}( \mathbf{c}_\alpha \leq \mathbf{R}_\alpha \boldsymbol{\eta} ) \prod_{i=1}^{N} \exp \left( -\xi \left( \mathbf{m}^\top_i \boldsymbol{\eta} \right) \right) }{ \prod_{i=1}^{N} \sup_{\boldsymbol{\eta}} \exp \left( -\xi \left( \mathbf{m}^\top_i \boldsymbol{\eta} \right) \right) } (2 \pi)^{-(p+2M+4)/2} \left| \mathbf{A}(\tau_0) \right|^{1/2} \\
& \geq
\exp \left[-\frac{1}{2} \left( \boldsymbol{\eta}^{\top} \mathbf{A}(\boldsymbol{\tau}) \boldsymbol{\eta} - 2  \boldsymbol{\eta}^{\top} \boldsymbol{\mu}_* \right) - \frac{1}{2} \boldsymbol{\mu}^{\top}_* \mathbf{A}(\boldsymbol{\tau})^{-1} \boldsymbol{\mu}_* \right] \\ 
& \times 
{ \mathbb{I}( \mathbf{c}_\alpha \leq \mathbf{R}_\alpha \boldsymbol{\eta} ) \prod_{i=1}^{N} \exp \left( -\xi \left( \mathbf{m}^\top_i \boldsymbol{\eta} \right) \right) } (2 \pi)^{-(p+2M+4)/2} \left| \mathbf{A}(\tau_0) \right|^{1/2}
\end{array}
$$
where $\mathbb{H}^{N}=[0,1]^N$ is a hypercube. 

Note that $
\exp \left[ - \frac{1}{2} \boldsymbol{\mu}^{\top}_* \mathbf{A}(\boldsymbol{\tau})^{-1} \boldsymbol{\mu}_* \right]
\geq 
\exp \left[ - \frac{1}{2} \boldsymbol{\mu}^{\top}_* \mathbf{A}({\tau_0})^{-1} \boldsymbol{\mu}_* \right] 
$. We have
$$
\begin{array}{rl}
& \displaystyle\int_{\mathbb{R}_{+}^2} \displaystyle\int_{\mathbb{H}^{N}} 
\pi( \boldsymbol{\eta} \mid \boldsymbol{\omega}, \boldsymbol{\tau}, \mathbf{y}) \pi\left(\boldsymbol{\omega} \mid \boldsymbol{\eta}^{\prime}, \mathbf{y}\right) 
d \boldsymbol{\omega} 
\pi\left(\boldsymbol{\tau} \mid \boldsymbol{\eta}^{\prime}, \mathbf{y}\right)  
d \boldsymbol{\tau} \\
& \geq 
\displaystyle\int_{\mathbb{R}_{+}^2} \Bigg\{
\exp \left[-\frac{1}{2} \boldsymbol{\beta}^{\top} \boldsymbol{\Sigma}^{-1} \boldsymbol{\beta} - \sum_{j \in \{ \alpha, \mathcal{B} \}} \frac{\tau_j}{2} \mathbf{u}_j^{\top} \mathbf{u}_j +\boldsymbol{\eta}^{\top} \boldsymbol{\mu}_* -\frac{1}{2} \boldsymbol{\mu}_*^{\top} \mathbf{A}\left(\tau_{0}\right)^{-1} \boldsymbol{\mu}_* \right] \\ 
& \times 
{ \mathbb{I}( \mathbf{c}_\alpha \leq \mathbf{R}_\alpha \boldsymbol{\eta} ) \prod_{i=1}^{N} \exp \left( -\xi \left( \mathbf{m}^\top_i \boldsymbol{\eta} \right) \right) } (2 \pi)^{-(p+2M+4)/2} \left| \mathbf{A}(\tau_0) \right|^{1/2} \\
& \times 
\prod_{j \in \{ \alpha, \mathcal{B} \}}
\tau_j^{a_{0}+ M/2} \exp \left[-\left(b_{0}+ {\mathbf{u}_j^{\prime}}^{\top} \mathbf{u}_j^{\prime} / 2\right) \tau_j \right] 
\frac{ \mathbb{I} \left(\tau_j \geq \tau_{0}\right) }{c_1\left(\tau_{0}, a_{0}+(M+2) / 2, b_{0}+{\mathbf{u}_j^{\prime}}^{\top} \mathbf{u}_j^{\prime} / 2\right)} 
\Bigg\} d \boldsymbol{\tau} \\
& \geq 
\exp \left[-\frac{1}{2} \boldsymbol{\beta}^{\top} \boldsymbol{\Sigma}^{-1} \boldsymbol{\beta} +\boldsymbol{\eta}^{\top} \boldsymbol{\mu}_* -\frac{1}{2} \boldsymbol{\mu}_*^{\top} \mathbf{A}\left(\tau_{0}\right)^{-1} \boldsymbol{\mu}_* \right] \\ 
& \times 
{ \mathbb{I}( \mathbf{c}_\alpha \leq \mathbf{R}_\alpha \boldsymbol{\eta} ) \prod_{i=1}^{N} \exp \left( -\xi \left( \mathbf{m}^\top_i \boldsymbol{\eta} \right) \right) } (2 \pi)^{-(p+2M+4)/2} \left| \mathbf{A}(\tau_0) \right|^{1/2} \\
& \times 
\displaystyle\int_{\mathbb{R}_{+}^2} \Bigg\{
\prod_{j \in \{ \alpha, \mathcal{B} \}}
\tau_j^{a_{0}+ M/2} \exp \left[-\left(b_{0} + {\mathbf{u}_j^{\prime}}^{\top} \mathbf{u}_j^{\prime} / 2 +{\mathbf{u}_j}^{\top} \mathbf{u}_j / 2 \right) \tau_j \right] \\
& \times
\frac{ \mathbb{I} \left(\tau_j \geq \tau_{0}\right) }{c_1\left(\tau_{0}, a_{0}+(M+2) / 2, b_{0} +{\mathbf{u}_j^{\prime}}^{\top} \mathbf{u}_j^{\prime} / 2\right)} 
\Bigg\} d \boldsymbol{\tau} .
\end{array}
$$
In addition to what has been noted in \cite{wang2018analysis}, we show with u-substitution, second fundamental theorem of calculus and chain rule or Leibniz integral rule, 
\begin{equation} \label{TG_inequality}
\begin{array}{rl}
&\frac{1}{c_1\left(\tau_{0}, a_{0}+ (M+2) / 2, b_{0}+ {\mathbf{u}_j^{\prime}}^{\top} \mathbf{u}_j^{\prime} / 2\right)}
\displaystyle\int_{\tau_{0}}^{\infty} \tau_j^{a_{0}+ M / 2} \exp \left[-\left(b_{0}+ {\mathbf{u}_j^{\prime}}^{\top} \mathbf{u}_j^{\prime} / 2+\mathbf{u}_j^{\top} \mathbf{u}_j / 2\right) \tau_j \right] d \tau_j \\
&=
\frac{
\left(b_{0}+ {\mathbf{u}_j^{\prime}}^{\top} \mathbf{u}_j^{\prime} / 2\right)^{a_{0}+ (M+2) / 2}
}{
\left(b_{0}+ {\mathbf{u}_j^{\prime}}^{\top} \mathbf{u}_j^{\prime} / 2+\mathbf{u}_j^{\top} \mathbf{u}_j / 2\right)^{a_{0}+ (M+2) / 2} 
}
\frac{
\displaystyle\int_{\left(b_{0}+ {\mathbf{u}_j^{\prime}}^{\top} \mathbf{u}_j^{\prime} / 2+\mathbf{u}_j^{\top} \mathbf{u}_j / 2\right) \tau_{0}}^\infty x^{a_{0}+ M/ 2} \exp (-x) d x
}{
\displaystyle\int_{\left(b_{0}+ {\mathbf{u}_j^{\prime}}^{\top} \mathbf{u}_j^{\prime} / 2\right) \tau_{0}}^{\infty} x^{a_{0}+ M / 2} \exp (-x) d x
} \\
& \geq 
\left(\frac{b_{0}}{b_{0}+\mathbf{u}_j^{\top} \mathbf{u}_j / 2}\right)^{a_{0}+ (M+2)/ 2}
\frac{
\displaystyle\int_{\left(b_{0}+ {\mathbf{u}_j^{\prime}}^{\top} \mathbf{u}_j^{\prime} / 2+\mathbf{u}_j^{\top} \mathbf{u}_j / 2\right) \tau_{0}}^\infty x^{a_{0}+ M/ 2} \exp (-x) d x
}{
\displaystyle\int_{\left(b_{0}+ {\mathbf{u}_j^{\prime}}^{\top} \mathbf{u}_j^{\prime} / 2\right) \tau_{0}}^{\infty} x^{a_{0}+ M / 2} \exp (-x) d x
} \\
& =
\left(\frac{b_{0}}{b_{0}+\mathbf{u}_j^{\top} \mathbf{u}_j / 2}\right)^{a_{0}+ (M+2)/ 2} 
\frac{
g_1({\mathbf{u}_j^{\prime}}^{\top} \mathbf{u}_j^{\prime} / 2)
}{
g_2({\mathbf{u}_j^{\prime}}^{\top} \mathbf{u}_j^{\prime} / 2)
} \\
& \geq
\left(\frac{b_{0}}{b_{0}+\mathbf{u}_j^{\top} \mathbf{u}_j / 2}\right)^{a_{0}+ (M+2)/ 2} \exp \left(-\tau_{0} \mathbf{u}_j^{\top} \mathbf{u}_j / 2\right).
\end{array}
\end{equation}
Here $g_3(v) = g_1(v) - \exp \left(-\tau_{0} \mathbf{u}_j^{\top} \mathbf{u}_j / 2\right) g_2(v)$, where $v = {\mathbf{u}_j^{\prime}}^{\top} \mathbf{u}_j^{\prime} / 2$ and 
$$
\begin{array}{rl}
\frac{d}{dv} g_3(v) =& 
- \left[ \left( \left( b_{0}+ {\mathbf{u}_j^{\prime}}^{\top} \mathbf{u}_j^{\prime} / 2+\mathbf{u}_j^{\top} \mathbf{u}_j / 2\right) \tau_{0} \right)^{a_{0}+ M/ 2} \exp \left( - \left(b_{0}+ {\mathbf{u}_j^{\prime}}^{\top} \mathbf{u}_j^{\prime} / 2+\mathbf{u}_j^{\top} \mathbf{u}_j / 2\right) \tau_{0} \right) \right] \tau_0 \\
& + \exp \left(-\tau_{0} \mathbf{u}_j^{\top} \mathbf{u}_j / 2\right) \left[ \left( \left(b_{0}+ {\mathbf{u}_j^{\prime}}^{\top} \mathbf{u}_j^{\prime} / 2\right) \tau_{0} \right)^{a_{0}+ M/ 2} \exp \left( -\left(b_{0}+ {\mathbf{u}_j^{\prime}}^{\top} \mathbf{u}_j^{\prime} / 2\right) \tau_{0} \right) \right] \tau_0 \\
=&
\left[
\left( \left(b_{0}+ {\mathbf{u}_j^{\prime}}^{\top} \mathbf{u}_j^{\prime} / 2\right) \tau_{0} \right)^{a_{0}+ M/ 2}
- \left( \left(b_{0}+ {\mathbf{u}_j^{\prime}}^{\top} \mathbf{u}_j^{\prime} / 2+\mathbf{u}_j^{\top} \mathbf{u}_j / 2\right) \tau_{0} \right)^{a_{0}+ M/ 2}
\right] \\
& \times
\exp \left( - \left(b_{0}+ {\mathbf{u}_j^{\prime}}^{\top} \mathbf{u}_j^{\prime} / 2+\mathbf{u}_j^{\top} \mathbf{u}_j / 2\right) \tau_{0} \right) \tau_0 \\
<& 0
\end{array}
$$
because of $\left[
\left( \left(b_{0}+ {\mathbf{u}_j^{\prime}}^{\top} \mathbf{u}_j^{\prime} / 2\right) \tau_{0} \right)^{a_{0}+ M/ 2}
- \left( \left(b_{0}+ {\mathbf{u}_j^{\prime}}^{\top} \mathbf{u}_j^{\prime} / 2+\mathbf{u}_j^{\top} \mathbf{u}_j / 2\right) \tau_{0} \right)^{a_{0}+ M/ 2}
\right] < 0$. We showed that $g_3(v)$ is a decreasing function. Together with the limit $\lim_{v \rightarrow \infty} g_3(v) = 0$, we see that $g_3(v) \geq 0$ and
$$
\begin{array}{rl}
g_3(v) &\geq 0 \\
g_1(v) &\geq \exp \left(-\tau_{0} \mathbf{u}_j^{\top} \mathbf{u}_j / 2\right) g_2(v) \\
g_1(v)/g_2(v) &\geq \exp \left(-\tau_{0} \mathbf{u}_j^{\top} \mathbf{u}_j / 2\right) \\
\frac{
\displaystyle\int_{\left(b_{0}+ {\mathbf{u}_j^{\prime}}^{\top} \mathbf{u}_j^{\prime} / 2+\mathbf{u}_j^{\top} \mathbf{u}_j / 2\right) \tau_{0}}^\infty x^{a_{0}+ M/ 2} \exp (-x) d x
}{
\displaystyle\int_{\left(b_{0}+ {\mathbf{u}_j^{\prime}}^{\top} \mathbf{u}_j^{\prime} / 2\right) \tau_{0}}^{\infty} x^{a_{0}+ M / 2} \exp (-x) d x
} 
&\geq \exp \left(-\tau_{0} \mathbf{u}_j^{\top} \mathbf{u}_j / 2\right) .
\end{array}
$$
Thus, from \eqref{TG_inequality} we have
$$
\begin{array}{rl}
k\left( \boldsymbol{\eta} \mid \boldsymbol{\eta}^{\prime}\right)
&\geq 
\exp \left[-\frac{1}{2} \boldsymbol{\beta}^{\top} \boldsymbol{\Sigma}^{-1} \boldsymbol{\beta} +\boldsymbol{\eta}^{\top} \boldsymbol{\mu}_* -\frac{1}{2} \boldsymbol{\mu}_*^{\top} \mathbf{A}\left(\tau_{0}\right)^{-1} \boldsymbol{\mu}_* \right] \\ 
& \times 
{ \mathbb{I}( \mathbf{c}_\alpha \leq \mathbf{R}_\alpha \boldsymbol{\eta} ) \prod_{i=1}^{N} \exp \left( -\xi \left( \mathbf{m}^\top_i \boldsymbol{\eta} \right) \right) } 
(2 \pi)^{-(p+2M+4)/2} \left| \mathbf{A}(\tau_0) \right|^{1/2} \\
& \times 
\prod_{j \in \{ \alpha, \mathcal{B} \}}
\left({b_{0}}/{(b_{0}+\mathbf{u}_j^{\top} \mathbf{u}_j / 2)}\right)^{a_{0}+ (M+2)/ 2} \exp \left(-\tau_{0} \mathbf{u}_j^{\top} \mathbf{u}_j / 2\right) \\
& \geq \delta h(\boldsymbol{\eta})
\end{array}
$$
where 
$$
\begin{array}{rl}
h( \boldsymbol{\eta} ) =& \exp \left[-\frac{1}{2} \boldsymbol{\beta}^{\top} \boldsymbol{\Sigma}^{-1} \boldsymbol{\beta} +\boldsymbol{\eta}^{\top} \boldsymbol{\mu}_* -\frac{1}{2} \boldsymbol{\mu}_*^{\top} \mathbf{A}\left(\tau_{0}\right)^{-1} \boldsymbol{\mu}_* \right] \\ 
& \times 
\frac{ \mathbb{I}( \mathbf{c}_\alpha \leq \mathbf{R}_\alpha \boldsymbol{\eta} ) }{c_{4}( \mathbf{M}, \mathbf{y})}
{ \prod_{i=1}^{N} \exp \left( -\xi \left( \mathbf{m}^\top_i \boldsymbol{\eta} \right) \right) } \\
& \times 
\prod_{j \in \{ \alpha, \mathcal{B} \}}
\left({b_{0}}/{(b_{0}+\mathbf{u}_j^{\top} \mathbf{u}_j / 2)}\right)^{a_{0}+ (M+2)/ 2} \exp \left(-\tau_{0} \mathbf{u}_j^{\top} \mathbf{u}_j / 2\right) \\
\delta = & (2 \pi)^{-(p+2M+4)/2} \left| \mathbf{A}(\tau_0) \right|^{1/2} { c_{4}( \mathbf{M}, \mathbf{y}) } < 1
\end{array}
$$
and
$$
\begin{array}{rl}
c_{4}( \mathbf{M}, \mathbf{y}) =& \exp \left[-\frac{1}{2} \boldsymbol{\mu}_*^{\top} \mathbf{A}\left(\tau_{0}\right)^{-1} \boldsymbol{\mu}_* \right] 
\oint_{ \mathbf{u}_\alpha \geq \mathbf{0} } 
\exp \left[-\frac{1}{2} \boldsymbol{\beta}^{\top} \boldsymbol{\Sigma}^{-1} \boldsymbol{\beta} + \boldsymbol{\eta}^{\top} \boldsymbol{\mu}_* \right] \\
& \times
{ \prod_{i=1}^{N} \exp \left( -\xi \left( \mathbf{m}^\top_i \boldsymbol{\eta} \right) \right) }\\
& \times 
\prod_{j \in \{ \alpha, \mathcal{B} \} }
\left(\frac{b_{0}}{b_{0}+\mathbf{u}_j^{\top} \mathbf{u}_j / 2}\right)^{a_{0}+ (M+2) / 2} \exp \left(-\tau_{0} \mathbf{u}_j^{\top} \mathbf{u}_j / 2\right) d \boldsymbol{\eta} \\
\leq &
\exp \left[-\frac{1}{2} \boldsymbol{\mu}_*^{\top} \mathbf{A}\left(\tau_{0}\right)^{-1} \boldsymbol{\mu}_* \right] 
\oint_{ \mathbf{u}_\alpha \geq \mathbf{0} } 
\exp \left[-\frac{1}{2} \boldsymbol{\beta}^{\top} \boldsymbol{\Sigma}^{-1} \boldsymbol{\beta} + \boldsymbol{\eta}^{\top} \boldsymbol{\mu}_* \right] \\
& \times 
\prod_{j \in \{\alpha, \mathcal{B} \} }
\exp \left(-\tau_{0} \mathbf{u}_j^{\top} \mathbf{u}_j / 2\right) d \boldsymbol{\eta} \\
=&
\oint_{ \mathbf{c}_\alpha \leq \mathbf{R}_\alpha \boldsymbol{\eta} } 
\exp \left[-\frac{1}{2} \left( \boldsymbol{\eta} - \mathbf{A}(\tau_0)^{-1} \boldsymbol{\mu}_* \right)^{\top} \mathbf{A}(\tau_0) \left( \boldsymbol{\eta} - \mathbf{A}(\tau_0)^{-1} \boldsymbol{\mu}_* \right) \right] d \boldsymbol{\eta} \\
<&
\int_{\mathbb{R}^{p+2M+4}}
\exp \left[-\frac{1}{2} \left( \boldsymbol{\eta} - \mathbf{A}(\tau_0)^{-1} \boldsymbol{\mu}_* \right)^{\top} \mathbf{A}(\tau_0) \left( \boldsymbol{\eta} - \mathbf{A}(\tau_0)^{-1} \boldsymbol{\mu}_* \right) \right] d \boldsymbol{\eta} \\
=& 
(2 \pi)^{(p+2M+4)/2}\left| \mathbf{A}\left(\tau_{0}\right)\right|^{-1 / 2} <\infty .
\end{array}
$$
This concludes the proof for uniform ergodicity. 
\end{proof}

\begin{proof}Proof of Theorem \ref{thm2}.
Recall that
$$
\int_{\mathbb{R}^{2}_+} \int_{\mathbb{H}^{N}} \pi(\boldsymbol{\eta}, \boldsymbol{\omega}, \boldsymbol{\tau} \mid \mathbf{y}) d \boldsymbol{\omega} d \boldsymbol{\tau} =\pi(\boldsymbol{\eta} \mid \mathbf{y} )
$$
and normalizing constant $c( \mathbf{y} )$,
$$
\begin{array}{c}
\pi(\boldsymbol{\eta}, \boldsymbol{\tau} \mid \mathbf{y}) = c( \mathbf{y} )^{-1} L(\boldsymbol{\eta} \mid \mathbf{y}, \mathbf{M} ) \pi (\boldsymbol{\eta} \mid \mathbf{b}, \mathbf{A}(\boldsymbol{\tau})^{-1}, \mathbf{R}_\alpha, \mathbf{c}_\alpha, \boldsymbol{\infty} ) 
\pi \left( \boldsymbol{\tau} \mid a_{0}, b_{0}, \tau_{0}\right)\\
c( \mathbf{y} ) = 
\displaystyle\int_{\mathbb{R}^{2}_+} \displaystyle\int_{\mathbb{R}^{p+2M+4}} 
{
L(\boldsymbol{\eta} \mid \mathbf{y}, \mathbf{M} )
\pi (\boldsymbol{\eta} \mid \mathbf{b}, \mathbf{A}(\boldsymbol{\tau})^{-1}, \mathbf{R}_\alpha, \mathbf{c}_\alpha, \boldsymbol{\infty} ) 
\pi \left( \boldsymbol{\tau} \mid a_{0}, b_{0}, \tau_{0}\right) } d \boldsymbol{\eta} d \boldsymbol{\tau} .
\end{array}
$$
Note that
$$
\begin{array}{rl}
\pi(\boldsymbol{\eta}, \boldsymbol{\omega}, \boldsymbol{\tau} \mid \mathbf{y} ) &= 
\pi(\boldsymbol{\eta}, \boldsymbol{\tau} \mid \mathbf{y}) \pi \left( \boldsymbol{\omega} \mid \boldsymbol{\eta} \right) \\
&= 
c( \mathbf{y} )^{-1} L(\boldsymbol{\eta} \mid \mathbf{y}, \mathbf{M} )
\pi (\boldsymbol{\eta} \mid \mathbf{b}, \mathbf{A}(\boldsymbol{\tau})^{-1}, \mathbf{R}_\alpha, \mathbf{c}_\alpha, \boldsymbol{\infty} ) 
\pi \left( \boldsymbol{\tau} \mid a_{0}, b_{0}, \tau_{0}\right) \pi \left( \boldsymbol{\omega} \mid \boldsymbol{\eta} \right).
\end{array}
$$
Thus, 
$$
\begin{array}{rl}
\pi(\boldsymbol{\eta}, \boldsymbol{\omega}, \boldsymbol{\tau} \mid \mathbf{y}) 
= &
c( \mathbf{y} )^{-1} c_1 \left(\tau_{0}, a_{0}, b_{0}\right)^{-2} |\mathbf{A}(\boldsymbol{\tau})|^{1/2} 2^{M+2} (2\pi)^{-(p+2M+4)/2} \mathbb{I} ( \mathbf{c}_\alpha \leq \mathbf{R}_\alpha \boldsymbol{\eta} ) \\
&\times
\exp \left\{\mathbf{y}^{\top} \mathbf{M} \boldsymbol{\eta}-
\frac{1}{2}\left(\boldsymbol{\eta}- \mathbf{b} \right)^{\top} 
\mathbf{A}(\boldsymbol{\tau})
\left(\boldsymbol{\eta}- \mathbf{b} \right) \right\} \\
&\times \prod_{j \in \{\alpha, \mathcal{B}\}} \tau_j^{a_{0}-1} e^{-b_{0} \tau_j} { \mathbb{I}\left(\tau_j \geq \tau_{0}\right) } \\
&\times 
\prod_{i=1}^{N} \left[ \exp \left(- \xi \left( \mathbf{m}^\top_i \boldsymbol{\eta} \right) \right) \pi \left( {\omega_i} \mid \boldsymbol{\eta} \right) \right] \\
=&
c( \mathbf{y} )^{-1} c_1 \left(\tau_{0}, a_{0}, b_{0}\right)^{-2} |\mathbf{A}(\boldsymbol{\tau})|^{1/2} 2^{M+2} (2\pi)^{-(p+2M+4)/2} \mathbb{I} ( \mathbf{c}_\alpha \leq \mathbf{R}_\alpha \boldsymbol{\eta} ) \\
&\times
\exp \left( \mathbf{y}^{\top} \mathbf{M} \boldsymbol{\eta} \right) \exp \left\{
- \frac{1}{2}\left(\boldsymbol{\eta}- \mathbf{b} \right)^{\top} 
\mathbf{A}(\boldsymbol{\tau})
\left(\boldsymbol{\eta}- \mathbf{b} \right) \right\} \\
&\times \prod_{j \in \{\alpha, \mathcal{B}\}} \tau_j^{a_{0}-1} e^{-b_{0} \tau_j} { \mathbb{I}\left(\tau_j \geq \tau_{0}\right) } \\
&\times 
\prod_{i=1}^{N} \left[ \exp \left(- \xi \left( \mathbf{m}^\top_i \boldsymbol{\eta} \right) \right) \pi \left( {\omega_i} \mid \boldsymbol{\eta} \right) \right] \\
=&
c( \mathbf{y} )^{-1} c_1 \left(\tau_{0}, a_{0}, b_{0}\right)^{-2} |\mathbf{A}(\boldsymbol{\tau})|^{1/2} 2^{M+2} (2\pi)^{-(p+2M+4)/2} \mathbb{I} ( \mathbf{c}_\alpha \leq \mathbf{R}_\alpha \boldsymbol{\eta} ) \\
&\times
\exp \left( \mathbf{y}^{\top} \mathbf{M} \boldsymbol{\eta} \right) \exp \left\{
- \frac{1}{2}\left(\boldsymbol{\eta}- \mathbf{b} \right)^{\top} 
\mathbf{A}(\boldsymbol{\tau})
\left(\boldsymbol{\eta}- \mathbf{b} \right) \right\} \\
&\times \prod_{j \in \{\alpha, \mathcal{B}\}} \tau_j^{a_{0}-1} e^{-b_{0} \tau_j} { \mathbb{I}\left(\tau_j \geq \tau_{0}\right) } \prod_{i=1}^{N} \pi\left( \omega_j \right) \mathbb{I} \left( \omega_j \leq \exp \left( -\xi \left(\mathbf{m}_{i}^{\top} \boldsymbol{\eta} \right) \right) \right)\\
\leq&
c( \mathbf{y} )^{-1} c_1 \left(\tau_{0}, a_{0}, b_{0}\right)^{-2} |\mathbf{A}(\boldsymbol{\tau})|^{1/2} 2^{M+2} (2\pi)^{-(p+2M+4)/2} \\
&\times
\exp \left( \mathbf{y}^{\top} \mathbf{M} \boldsymbol{\eta} \right) \exp \left\{
- \frac{1}{2}\left(\boldsymbol{\eta}- \mathbf{b} \right)^{\top} 
\mathbf{A}(\boldsymbol{\tau})
\left(\boldsymbol{\eta}- \mathbf{b} \right) \right\} \\
&\times \prod_{j \in \{\alpha, \mathcal{B}\}} \tau_j^{a_{0}-1} e^{-b_{0} \tau_j} { \mathbb{I}\left(\tau_j \geq \tau_{0}\right) } \pi (\boldsymbol{\omega}) \prod_{i=1}^{N} \mathbb{I} \left( \omega_j \leq \exp \left( -\xi \left(\mathbf{m}_{i}^{\top} \boldsymbol{\eta} \right) \right) \right) \\
=&
c( \mathbf{y} )^{-1} c_1 \left(\tau_{0}, a_{0}, b_{0}\right)^{-2} 2^{M+2} \exp \left( \mathbf{y}^{\top} \mathbf{M} \boldsymbol{\eta} \right) \pi(\boldsymbol{\eta} | \mathbf{b}, \mathbf{A}(\boldsymbol{\tau})^{-1} ) \\
&\times \prod_{j \in \{\alpha, \mathcal{B}\}} \tau_j^{a_{0}-1} e^{-b_{0} \tau_j} { \mathbb{I}\left(\tau_j \geq \tau_{0}\right) } \pi (\boldsymbol{\omega}) \prod_{i=1}^{N} \mathbb{I} \left( \omega_j \leq \exp \left( -\xi \left(\mathbf{m}_{i}^{\top} \boldsymbol{\eta} \right) \right) \right) 
\end{array}
$$
where $\pi(\boldsymbol{\eta} | \mathbf{b}, \mathbf{A}(\boldsymbol{\tau})^{-1} )$ is the PDF of $\boldsymbol{\eta} \sim \mathrm{N} (\mathbf{b}, \mathbf{A}(\boldsymbol{\tau})^{-1})$. We can integrate out $\boldsymbol{\omega}$ and upper bound using $\exp \left( -\xi \left(\mathbf{m}_{i}^{\top} \boldsymbol{\eta} \right) \right) \leq 1$,
$$
\begin{array}{rl}
\int_{\mathbb{H}^{N}} \pi(\boldsymbol{\eta}, \boldsymbol{\omega}, \boldsymbol{\tau} \mid \mathbf{y}) d \boldsymbol{\omega} 
=& 
\pi(\boldsymbol{\eta}, \boldsymbol{\tau} \mid \mathbf{y}) \\
\leq&
c( \mathbf{y} )^{-1} c_1 \left(\tau_{0}, a_{0}, b_{0}\right)^{-2} 2^{M+2} \exp \left( \mathbf{y}^{\top} \mathbf{M} \boldsymbol{\eta} \right) \pi(\boldsymbol{\eta} | \mathbf{b}, \mathbf{A}(\boldsymbol{\tau})^{-1} ) \\
&\times \prod_{j \in \{\alpha, \mathcal{B}\}} \tau_j^{a_{0}-1} e^{-b_{0} \tau_j} { \mathbb{I}\left(\tau_j \geq \tau_{0}\right) } \\
&\times \prod_{i=1}^{N} \exp \left( -\xi \left(\mathbf{m}_{i}^{\top} \boldsymbol{\eta} \right) \right) \\
\leq&
c( \mathbf{y} )^{-1} c_1 \left(\tau_{0}, a_{0}, b_{0}\right)^{-2} 2^{M+2} \exp \left( \mathbf{y}^{\top} \mathbf{M} \boldsymbol{\eta} \right) \pi(\boldsymbol{\eta} | \mathbf{b}, \mathbf{A}(\boldsymbol{\tau})^{-1} ) \\
&\times \prod_{j \in \{\alpha, \mathcal{B}\}} \tau_j^{a_{0}-1} e^{-b_{0} \tau_j} { \mathbb{I}\left(\tau_j \geq \tau_{0}\right) } .
\end{array}
$$
Here we set $\mathbf{z} = \mathbf{t} + \mathbf{M}^\top \mathbf{y}$ in order to facilitate the proof, where $\int \exp( \boldsymbol{\eta}^{\top} \mathbf{z} ) \pi(\boldsymbol{\eta} | \mathbf{b}, \mathbf{A}(\boldsymbol{\tau})^{-1} ) d\boldsymbol{\eta} $ is of the form of a normal MGF.
We have
$$
\begin{array}{rl}
\int_{\mathbb{R}^{2}_+} 
\int_{\mathbb{R}^{p+2M+4}} e^{\boldsymbol{\eta}^{\top} \mathbf{t} }
\pi(\boldsymbol{\eta}, \boldsymbol{\tau} \mid \mathbf{y})
d \boldsymbol{\eta} d \boldsymbol{\tau} 
\leq &
\displaystyle\int_{\mathbb{R}^{2}_+} 
\displaystyle\int_{\mathbb{R}^{p+2M+4}} \Bigg\{
c( \mathbf{y} )^{-1} c_1 \left(\tau_{0}, a_{0}, b_{0}\right)^{-2} 2^{M+2} \\
&\times \exp( \boldsymbol{\eta}^{\top} \mathbf{t} )  \exp \left( \mathbf{y}^{\top} \mathbf{M} \boldsymbol{\eta} \right) \\
&\times \pi(\boldsymbol{\eta} | \mathbf{b}, \mathbf{A}(\boldsymbol{\tau})^{-1} ) \prod_{j \in \{\alpha, \mathcal{B}\}} \tau_j^{a_{0}-1} e^{-b_{0} \tau_j} \mathbb{I}\left(\tau_j \geq \tau_{0}\right)
\Bigg\} d \boldsymbol{\eta} d \boldsymbol{\tau} \\
= &
\displaystyle\int_{\mathbb{R}^{2}_+} 
\displaystyle\int_{\mathbb{R}^{p+2M+4}} \Bigg\{
c( \mathbf{y} )^{-1} c_1 \left(\tau_{0}, a_{0}, b_{0}\right)^{-2} 2^{M+2} \exp( \boldsymbol{\eta}^{\top} \mathbf{z} ) \\ 
&\times \pi(\boldsymbol{\eta} | \mathbf{b}, \mathbf{A}(\boldsymbol{\tau})^{-1} ) \prod_{j \in \{\alpha, \mathcal{B}\}} \tau_j^{a_{0}-1} e^{-b_{0} \tau_j} { \mathbb{I}\left(\tau_j \geq \tau_{0}\right) }
\Bigg\} d \boldsymbol{\eta} d \boldsymbol{\tau} \\
= &
c( \mathbf{y} )^{-1} c_1 \left(\tau_{0}, a_{0}, b_{0}\right)^{-2} 2^{M+2} \\
&\times \displaystyle\int_{\mathbb{R}^{2}_+} \Bigg\{
\exp \left( \mathbf{b}^{\top} \mathbf{z} + \frac{1}{2} \mathbf{z}^{\top} \mathbf{A}(\boldsymbol{\tau})^{-1} \mathbf{z} \right) \\
&\times 
\prod_{j \in \{\alpha, \mathcal{B}\}} \tau_j^{a_{0}-1} e^{-b_{0} \tau_j} { \mathbb{I}\left(\tau_j \geq \tau_{0}\right) } 
\Bigg\} d \boldsymbol{\tau} \\
\leq &
c( \mathbf{y} )^{-1} c_1 \left(\tau_{0}, a_{0}, b_{0}\right)^{-2} 2^{M+2}
\exp \left( \mathbf{b}^{\top} \mathbf{z} + \frac{1}{2} \mathbf{z}^{\top} \mathbf{A}({\tau_0})^{-1} \mathbf{z} \right) \\
&\times 
\displaystyle\int_{\mathbb{R}^{2}_+} \Bigg\{
\prod_{j \in \{\alpha, \mathcal{B}\}} \tau_j^{a_{0}-1} e^{-b_{0} \tau_j} { \mathbb{I}\left(\tau_j \geq \tau_{0}\right) } 
\Bigg\} d \boldsymbol{\tau}\\
< & \infty
\end{array}
$$
and the moment generating function exist. This concludes the proof.
\end{proof}


\backmatter




%




\bibliographystyle{biom} 
\bibliography{references.bib}


\end{document}